\documentstyle[12pt]{article}                    
\newfont{\ffont}{msym10}                          
\newcommand{\beq}{\begin{equation}}               
\newcommand{\eeq}{\end{equation}}                 
\newcommand{\bqry}{\begin{eqnarray}}              
\newcommand{\eqry}{\end{eqnarray}}                
\newcommand{\bqryn}{\begin{eqnarray*}}            
\newcommand{\eqryn}{\end{eqnarray*}}              
\newcommand{\NL}{\nonumber \\}                    
\newcommand{\preprint}[1]{\begin{table}[t]        
            \begin{flushright}                    
            \begin{large}{#1}\end{large}          
            \end{flushright}                      
            \end{table}}                          
\newcommand{\PD}[2]                               
    {\frac{\partial^{#2}}{\partial #1^{#2}}}      
\begin{document}
\preprint{LA-UR-97-3795-rev/short}
\title{New Mass and Mass-Mixing Angle Relations \\ for Pseudoscalar Mesons}
\author{\\ L. Burakovsky\thanks{E-mail: BURAKOV@T5.LANL.GOV} \
and \ T. Goldman\thanks{E-mail: GOLDMAN@T5.LANL.GOV} \
\\  \\  Theoretical Division, MS B285 \\ Los Alamos National Laboratory \\ 
Los Alamos, NM 87545, USA \\}
\date{\today }
\maketitle
\begin{abstract}
We study the origins of the inaccuracies of Schwinger's nonet mass, and the 
Sakurai mass-mixing angle, formulae for the pseudoscalar meson nonet, and 
suggest new versions of them, modified by the inclusion of the pseudoscalar 
decay constants. We use these new formulae to determine the pseudoscalar decay
constants and mixing angle. The results obtained, $f_8/f_\pi =
1.185\pm 0.040,$ $f_9/f_\pi =1.095\pm 0.020,$ $f_\eta /f_\pi =1.085\pm 
0.025,$ $f_{\eta ^{'}}/f_\pi =1.195\pm 0.035,$ $\theta =
(-21.4\pm 1.0)^o,$ are in excellent agreement with experiment.   
\end{abstract}
\hspace*{0.85in}mesons \\

\newpage
\section{Introduction }
Schwinger's original nonet mass formula \cite{Sch} (here the symbol for the
meson stands either for its mass {\it or} mass squared),
\beq
(4K-3\eta -\pi )(3\eta ^{'}+\pi -4K)=8(K-\pi ) ^2,
\eeq
and the Sakurai mass-mixing angle formula \cite{Sakurai},
\beq
\tan ^2\theta =\frac{4K-3\eta -\pi }{3\eta ^{'}+\pi -4K},
\eeq
both relate the masses of the isovector $(\pi ),$ isodoublet $(K)$ and 
isoscalar mostly octet $(\eta )$ and mostly singlet $(\eta ^{'})$ states of a 
meson nonet, and the nonet mixing angle $(\theta ).$ We alert the reader that,
although we use notation suggestive of masses below, each formula is to be 
reprised in terms of mass or mass squared values in this introductory section.

The relations (1) and (2) are usually derived in the following way:
For a meson nonet, the isoscalar octet-singlet mass matrix,
\beq
{\cal M}=\left( 
\begin{array}{cc}
M_{88} & M_{89} \\
M_{89} & M_{99}
\end{array}
\right ) ,
\eeq 
is diagonalized by the masses of the physical $\eta $ and $\eta ^{'}$ states:
\bqry
{\cal M} & = & \left( 
\begin{array}{rr}
\cos \theta & \sin \theta \\
-\sin \theta & \cos \theta
\end{array}
\right) \left(
\begin{array}{cc}
\eta  & 0  \\
0 & \eta ^{'}
\end{array}
\right) \left(
\begin{array}{rr}
\cos \theta & -\sin \theta \\
\sin \theta & \cos \theta
\end{array}
\right)   \NL
 & = & \left(
\begin{array}{cc}
\cos ^2\theta \;\eta +\sin ^2\theta \;\eta ^{'} & \sin \theta \cos \theta \;\!
(\eta ^{'}-\eta ) \\
\sin \theta \cos \theta \;\!(\eta ^{'}-\eta )  & \sin ^2\theta \;\eta +\cos ^2
\theta \;\eta ^{'}
\end{array}
\right) ,
\eqry
where $\theta $ is the nonet mixing angle, which is determined by comparing 
the corresponding quadrants of the matrices (3) and (4), by any of the three
following relations: 
\beq
\tan ^2\theta =\frac{M_{88}-\eta }{\eta ^{'}-M_{88}},
\eeq
\beq
\tan ^2\theta =\frac{\eta ^{'}-M_{99}}{M_{99}-\eta },
\eeq
\beq
\sin 2\theta =\frac{2M_{89}}{\eta ^{'}-\eta }.
\eeq
It is easily seen that Eqs. (5) and (6) are identical, since, due to the trace
invariance of ${\cal M},$
\beq
\eta +\eta ^{'}=M_{88}+M_{99},
\eeq
and therefore, $M_{88}-\eta =\eta ^{'}-M_{99},$ and $\eta ^{'}-M_{88}=
M_{99}-\eta .$ Eliminating $\theta $ from (5),(7), or (6),(7), with the 
help of $\sin 2\theta =2\tan \theta /(1+\tan ^2\theta ),$ leads, respectively,
to
\beq
(\eta -M_{88})(M_{88}-\eta ^{'})=M_{89}^2,
\eeq
\beq
(M_{99}-\eta )(\eta ^{'}-M_{99})=M_{89}^2,
\eeq
which again are identical, through (8). 

We note that, under the quark model inspired conditions
\beq
M_{88}=\frac{4K-\pi }{3},\;\;\;M_{99}=\frac{2K+\pi }{3},\;\;\;M_{89}=-\frac{
2\sqrt{2}}{3}(K-\pi ),
\eeq
where the first of the three relations in (11) is the standard
Gell-Mann--Okubo mass formula \cite{GMO}, Eqs. (9),(10) have the
effectively unique solution
\bqry
\eta & = & 2K-\pi ,\;\eta ^{'}\;=\;\pi ,\;\theta \;=\;\arctan \frac{1}{\sqrt{
2}}\;\cong \;35.3^o,\;\;{\rm or}  \\
\eta & = & \pi ,\;\eta ^{'}\;=\;2K-\pi ,\;\theta \;=\;-\arctan \sqrt{2}\cong 
-54.7^o,
\eqry
which corresponds to the ``ideal'' structure of a meson nonet.

For all well established meson nonets, except the pseudoscalar (and, we expect,
scalar) one(s), both linear and quadratic versions of Eqs. (5)-(10) are in good
agreement with experiment. For example, for vector mesons, if one assumes the 
validity of the Gell-Mann--Okubo formula $\omega _8=(4K^\ast -\rho )/3,$ then
one obtains from Eq. (9) with the measured meson masses \cite{pdg}, $M_{89}=
-0.209\pm 0.001$ GeV$^2$ in the quadratic case, and $-0.113\pm 0.001$ GeV in 
the linear case. Note that this is entirely consistent with $-0.196\pm 0.005$ 
GeV$^2$ and $-0.118\pm 0.003$ GeV, respectively, which follow from the third 
element of (16). For tensor mesons, a similar comparison gives 
$-0.305\pm 0.020$ GeV$^2$ vs. $-0.287\pm 0.015$ GeV$^2,$ and $-0.105\pm 0.008$
GeV vs. $-0.104\pm 0.005$ GeV, respectively.  

However, for the pseudoscalar nonet, one obtains from Eq. (5) with meson masses
squared, $\theta \approx -11^o,$ in sharp disagreement with 
experiment, which favors the $\eta $-$\eta ^{'}$ mixing angle in the vicinity 
of $-20^o$ \cite{pdg,data,Abele}. Although using linear meson masses in Eq. (5)
does give $\theta \approx -24^o,$ in better agreement with 
data than its mass-squared counterpart, the value of $M_{89},$ as given by (9),
is now $-0.165\pm 0.004$ GeV, vs. $-2\sqrt{2}/3\;(K-\pi )=-0.338\pm 0.004$ GeV.
This emphasizes that neither Schwinger's nonet mass formula nor the mass-mixing
angle relations (including Sakurai's) (5)-(7) hold for the pseudoscalar nonet. 

It is well known, however, that the pseudoscalar (and, probably, scalar) mass 
spectrum does not follow the ``ideal'' structure, Eq. (11), since the mass of 
the pseudoscalar isoscalar singlet state is shifted up from its ``ideal'' 
value of $(2K+\pi )/3,$ presumably by the instanton-induced 't Hooft 
interaction \cite{tH} which breaks axial U(1) symmetry 
\cite{symbr,Dmitra,Dmitra2}. However, the use of $M_{99}=(2K+\pi )/3+A,$ $A\neq
0,$ in Eqs. (5)-(8) will again lead to Schwinger's formula (9), which does not
hold for the pseudoscalar mesons, as just demonstrated. [In fact, the 
structure of this formula does not depend at all on $M_{99},$ as seen in (9).]
Therefore, instanton, as well as any other effects which may shift the mass 
of the pseudoscalar isoscalar singlet state, cannot constitute the explanation
of the failure of Schwinger's quartic mass and the Sakurai mass-mixing angle 
formulae for the pseudoscalar nonet. We believe, however, that the following
analysis can resolve this problem.

\section{Pseudoscalar meson mass squared matrix}
Here we suggest the following form of the mass squared matrix for 
the pseudoscalar mesons:
\beq
\bar{f}^2M^2=\frac{1}{3}\left(
\begin{array}{cc}
4f_K^2K^2-f_\pi ^2\pi ^2 & -2\sqrt{2}\;(f_K^2K^2-f_\pi ^2\pi ^2) \\
-2\sqrt{2}\;(f_K^2K^2-f_\pi ^2\pi ^2) & 2f_K^2K^2+f_\pi ^2\pi ^2+3f_9^2A 
\end{array}
\right) ,
\eeq
where $f$'s are the pseudoscalar decay constants defined below, and $A$ stands
for the sum of all possible contributions to the shift of the isoscalar
singlet mass (from instanton effects, $1/N_c$-expansion diagrams, gluon
annihilation diagrams, etc.). 

Indeed, the form of such a mass squared matrix is determined by the form of
Gell-Mann--Okubo type relations among the masses of the isovector, isodoublet,
and isoscalar octet and singlet states (which in our case are Eqs. (17)-(19)
below), since this matrix must be equivalent to that of the form (3), which in
the case we are considering is
\beq
\bar{f}^2M^2=\left(
\begin{array}{cc}
f_8^2\eta _{88}^2 & f_8f_9\eta _{89}^2 \\
f_8f_9\eta _{89}^2 & f_9^2\eta _{99}^2
\end{array}
\right) ,
\eeq
and which we assume is diagonal in the basis of the physical $\eta $
and $\eta ^{'}$ meson masses and decay constants:
\beq
\bar{f}^2M^2=\left(
\begin{array}{cc}
f_\eta ^2\eta ^2 & 0  \\
0 & f_{\eta ^{'}}^2\eta ^{'2}
\end{array}
\right) .
\eeq
The equivalence of the matrices (14) and (16) is guaranteed by the validity of
the following relations: 
\beq
f_8^2\eta _{88}^2=-\frac{1}{3}\left[ m_u\langle \bar{u}u\rangle +m_d\langle 
\bar{d}d\rangle +4m_s\langle \bar{s}s\rangle \right] =\frac{4f_K^2K^2-f_\pi ^2
\pi ^2}{3}, 
\eeq
\beq
f_8f_9\eta _{89}^2=-\frac{\sqrt{2}}{3}\left[ m_u\langle \bar{u}u\rangle +m_d
\langle \bar{d}d\rangle -2m_s\langle \bar{s}s\rangle \right] =\frac{2\sqrt{
2}}{3}\left( f_\pi ^2\pi ^2-f_K^2K^2\right) ,
\eeq
\beq
f_9^2\eta _{99}^2=f_9^2A-\frac{2}{3}\left[ m_u\langle \bar{u}u\rangle +m_d
\langle \bar{d}d\rangle +m_s\langle \bar{s}s\rangle \right] =f_9^2A\;+\;\frac{
2f_K^2K^2+f_\pi ^2\pi ^2}{3},
\eeq 
with 
\beq
K^2\equiv \frac{(K^\pm )^2+(K^0)^2}{2},
\eeq
as suggested by Dmitrasinovic \cite{Dmitra2}, on the basis of the (precise)
Gell-Mann--Oakes-Renner formulae which relate the pseudoscalar
masses and decay constants to the quark masses and condensates \cite{GOR}:
\bqry
f_\pi ^2\;\pi ^2 & = & -\left[ m_u\langle \bar{u}u\rangle +m_d\langle \bar{d}d
\rangle \right] , \\
f_K^2\left( K^\pm \right) ^2 & = & -\left[ m_u\langle \bar{u}u\rangle +m_s
\langle \bar{s}s\rangle \right] , \\
f_K^2\left( K^0\right) ^2 & = & -\left[ m_d\langle \bar{d}d\rangle +m_s\langle
\bar{s}s\rangle \right] .
\eqry
(We ignore Dashen's theorem violating effects \cite{Dash}, and only 
approximately take into account isospin violating effects via (20), as we are
not concerned here with accuracies better than 1\%.)
 
Note that in the limit of exact nonet symmetry,
\beq
f_\pi =f_K=f_8=f_9\equiv \bar{f},
\eeq
the mass squared matrix (14) reduces to (3) with the matrix elements defined
in (11) and an additional contribution in the isoscalar singlet channel $(A).$

\section{Modified Gell-Mann--Okubo mass formula}
Here, we shall only explicitly demonstrate the validity of the modified 
Gell-Mann--Okubo formula (17) (the remaining relations (18) and (19) may be 
checked in a similar way). This formula may be obtained in standard chiral 
perturbation theory \cite{GL}, as follows.

Standard chiral perturbation theory leads to the following expressions for
the pseudoscalar meson masses and decay constants~\cite{GL}:
\bqry
\pi ^2 & = & 2m_nB\left( 1+\mu _\pi -\frac{1}{3}\mu 
_8+2m_nK_3+(2m_n+m_s)K_4\right) , \NL
K^2 & = & (m_n+m_s)B\left( 1+\frac{2}{3}\mu _8+(m_n+m_s)K_3 
+(2m_n+m_s)K_4\right) 
, \NL
\eta _{88}^2 & = & \frac{2}{3}(m_n+2m_s)B\left( 1+2\mu _K-\frac{4}{3}\mu _8
+\frac{2}{3}(m_n+2m_s)K_3+(2m_n+m_s)K_4\right) \NL
 & + & 2m_nB\left( -\mu _\pi+\frac{2}{3}\mu _K+\frac{1}{3}\mu _8\right) +
B(m_s-m_n)^{2}K_5, \NL
f_\pi & = & \bar{f}\Big( 1-2\mu _\pi-\mu _K+2m_nK_6+(m_n+2m_s)K_7\Big) , \NL
f_K & = & \bar{f}\left( 1-\frac{3}{4}\mu _\pi -\frac{3}{2}\mu _K-\frac{3}{4}
\mu _8+(m_n+m_s)K_6+(m_n+2m_s)K_7\right) , \NL
f_8 & = & \bar{f}\left( 1-3\mu _K+\frac{2}{3}(m_n+2m_s)K_6 
+(m_n+2m_s)K_7\right) ,
\eqry
where $\mu $'s are chiral logarithms, and the constants $K_i$
are\footnote{We have extracted factors absorbed into $K_{4}, K_{5}$ and
$K_{7}$ in ref. \cite{GL}, to make uniformly explicit the quark mass
factors and the overall scale.} proper combinations of the low energy
coupling constants $L_i.$ It then follows from these relations that the
standard Gell-Mann--Okubo formula is broken in first nonleading
order~\cite{GL},
\bqry
\triangle _{{\rm GMO}} & \equiv  & 4K^2 -3\eta_{88}^2 -\pi^2 \NL
 & = & 4B\Big[ m_n(\mu _\pi +\mu _8-2\mu _K)+2m_s(\mu _8-\mu _K) \Big] +\ldots
\;,
\eqry
where $\ldots $ indicates higher order terms. 

However, the modified Gell-Mann--Okubo formula remains valid in this
order, and is violated only by second (non-leading) order SU(3)-flavor
breaking effects:
\bqry
\triangle ^{'}_{{\rm GMO}} & \equiv & \frac{1}{{\bar f}^2}\left( 4f_K^2K^2-
3f_8^2\eta _{88}^2-f_\pi ^2\pi ^2\right)  \NL
 & = & 4B\;(m_s-m_n) \left[ \left( \mu_K+\frac{1}{2}\mu_8 -\frac{3}{2}\mu_\pi 
\right) \right.  \NL
 & &  \left. -\;(m_s-m_n)\left(\frac{1}{3}K_{3}+\frac{3}{4}K_{5} 
+\frac{2}{3}K_{7}\right) \right] +\ldots \;,
\eqry
since the second factor in the first bracketed term on the r.h.s. of
(27) must vanish in the SU(3)-flavor limit. 

This analysis shows, in an essentially model independent way, that the
modified Gell-Mann--Okubo formula (17) must be a more accurate relation
among the octet pseudoscalar mesons than the standard one,\footnote{A
search for a relation of a more general form, $4f_K^aK^2=3f_8^a\eta
_{88}^2+f_\pi ^a\pi ^2,$ which would hold in the first nonleading order
of standard chiral perturbation theory, results in $a= 2.$} and
therefore, use of the form of the mass squared matrix (14) is fully
justified.\footnote{It has been suggested in the literature that the
pseudoscalar decay constants should enter relations like (17)-(19) in
the first rather than second power \cite{first}. As discussed above,
such relations are expected to be less accurate than ours, according to
chiral perturbation theory.}

\section{The Schwinger and Sakurai formulae reexamined}
Starting with the mass squared matrix (14), the considerations which lead to
Eqs. (5)-(10) above, will now lead, through (17)-(19), to the following two 
relations,
\beq
\tan ^2\theta =\frac{4f_K^2K^2-3f_\eta ^2\eta ^2-f_\pi ^2\pi 
^2}{3f_{\eta ^{'}}^2\eta ^{'2}+f_\pi ^2\pi ^2-4f_K^2K^2},
\eeq
\beq
\left( 4f_K^2K^2-3f_\eta ^2\eta ^2-f_\pi ^2\pi ^2\right) \left( 3f_{\eta ^{
'}}^2\eta ^{'2}+f_\pi ^2\pi ^2-4f_K^2K^2\right) =8\left( f_K^2K^2-f_\pi ^2\pi 
^2\right) ^2,
\eeq
where $\theta$ is the mixing angle for the valence quark wavefunctions
which produces the mass eigenstates from the octet-singlet basis, and
which we refer to as ``the Schwinger nonet mass, and Sakurai
mass-mixing angle (respectively), formulae reexamined''.

In contrast to $f_\pi $ and $f_K,$ the values of which are well
established experimentally \cite{pdg},
\beq
\sqrt{2}f_K=159.8\pm 1.6\;{\rm MeV,}\;\;\;\sqrt{2}f_\pi=130.7\pm 0.3\;{\rm 
MeV},\;\;\;\frac{f_K}{f_\pi }=1.22\pm 0.01
\eeq
the values of $f_\eta ,$ $f_{\eta ^{'}}$ and $\theta $ are known rather
poorly. We now wish to calculate these values using the relations
(28),(29), and compare the results with available experimental data. It
is obvious that the two relations are not enough for determining the
three unknowns. However, the additional relation, independent of
(28),(29) (the trace condition for (14),(16)),
\beq
f_\eta ^2\eta ^2+f_{\eta ^{'}}^2\eta ^{'2}=2f_K^2K^2+f_9^2A,
\eeq
introduces yet an additional unknown, $A$. We therefore develop another
independent relation among $f_\eta ,$ $f_{\eta ^{'}}$ and $\theta$, as
follows. 

The light neutral pseudoscalar decay constants are defined by the
matrix elements
\beq
\langle 0| \sum_q c^P_{q} \bar{\psi}_{q}(0) \gamma^\mu \gamma^5 
\psi_{q}(0) |P(p) \rangle = i \delta^{jP} f_P p^\mu ,
\eeq
where $\psi_{q}$, $q=(u,d,s)$ are the quark field operators, and we
define the wave function of a neutral pseudoscalar meson $P$ in terms
of the quark basis states $q\bar{q}$ as:
\beq
|P\rangle = \sum _q c^P_q |q\bar{q}\rangle ,\;\;\;q=u,d,s,
\eeq
where for $P = \pi^0$, $c^3_u=1/\sqrt{2} = -c^3_d$, $c^3_s=0$, for $P =
\eta_{88}$, $c^8_u = c^8_d = 1/\sqrt{6}$, $c^8_s = -2/\sqrt{6}$ and for
$P = \eta_{99}$, $c^9_u = c^9_d = c^9_s = 1/\sqrt{3}$.

The pseudoscalar decay constants defined in (32) can now be expressed as
\beq
f_P = \sum _q(c^P_{q})^{2}f_{q\bar{q}},
\eeq
where we have introduced the auxiliary decay constants $f_{q\bar{q}}$ defined 
as the decay constants of the $q\bar{q}$ pseudoscalar bound states having the
masses $M(q\bar{q}).$ In the isospin limit, $f_{u\bar{u}}=f_{d\bar{d}}=f_{u
\bar{d}}=f_{\pi ^0}=f_{\pi ^{+}}.$ Using this approximation, and evaluating 
the appropriate matrix elements leads to the following relations:
\bqry
f_\eta  & = & \left( \frac{\cos \theta -\sqrt{2}\sin \theta }{\sqrt{3}}\right)
^2f_\pi \;+\;\left( \frac{\sin \theta +\sqrt{2}\cos \theta }{\sqrt{3}}\right)
^2f_{s\bar{s}}, \\
f_{\eta ^{'}} & = & \left( \frac{\sin \theta +\sqrt{2}\cos \theta }{\sqrt{3}}
\right) ^2f_\pi \;+\;\left( \frac{\cos \theta -\sqrt{2}\sin \theta }{\sqrt{3}}
\right) ^2f_{s\bar{s}}.
\eqry  

Now we have four equations, (28),(29),(35),(36), which allow us to
determine the three unknowns, $f_\eta ,$ $f_{\eta ^{'}},$ $\theta ,$ as
well as the additional quantity introduced, namely, $f_{s\bar{s}}.$ The
solution to these four equations is
\bqry
\frac{f_\eta }{f_\pi } & = & 1.085\pm 0.025, \\
\frac{f_{\eta ^{'}}}{f_\pi } & = & 1.195\pm 0.035, \\
\frac{f_{s\bar{s}}}{f_\pi } & = & 1.280\pm 0.060, \\
\theta  & = & (-21.4\pm 1.0)^o.
\eqry
[The $\pi $ and $K$ electromagnetic mass differences, and the uncertainties in
the values of $f_\pi $ and $f_K,$ (see (30)), are taken as a measure of 
the uncertainties of the results.] 

Before comparing the solution obtained with
experiment, let us also calculate the values of $f_8$ and $f_9$ which are 
obtained from (35),(36) in the no-mixing case $(\theta =0):$
\bqry 
f_8 & = & \frac{1}{3}f_\pi \;+\;\frac{2}{3}f_{s\bar{s}}, \\
f_9 & = & \frac{2}{3}f_\pi \;+\;\frac{1}{3}f_{s\bar{s}}.
\eqry
Therefore, as follows from (39),(41),(42),
\bqry
\frac{f_8}{f_\pi } & = & 1.185\pm 0.040, \\
\frac{f_9}{f_\pi } & = & 1.095\pm 0.020.
\eqry

The $\eta $-$\eta ^{'}$ mixing angle, as given in (40), is in agreement with
most of experimental data which concentrate around $-20^o$ 
\cite{pdg,data,Abele}. Also, the values for $f_8/f_\pi,$ $f_9/f_\pi $ and 
$\theta $ are consistent with those suggested in the literature, as we show in
Table I.

\begin{center}
\begin{tabular}{|c|c|c|c|c|} \hline
 Ref. & $f_8/f_\pi $ & $f_9/f_\pi $ & $\theta ,$ deg.  \\ 
\hline
 This work & $1.185\pm 0.040$ & $1.095\pm 0.020$ & $-21.4\pm 1.0$  \\ \hline
    [6]    & $1.11\pm 0.06$ & $1.10\pm 0.02$ & $-16.4\pm 1.2$   \\ \hline
  [15,16]  &     $1.25$     & $1.04\pm 0.04$ &   $-23\pm 3$     \\ \hline
    [17]   & $1.33\pm 0.02$ & $1.05\pm 0.04$ &   $-22\pm 3$     \\ \hline
    [18]   & $1.12\pm 0.14$ & $1.04\pm 0.08$ & $-18.9\pm 2.0$   \\ \hline
    [19]   & $1.38\pm 0.22$ & $1.06\pm 0.03$ & $-22.0\pm 3.3$   \\ \hline
    [20]   &     1.254      &     1.127      &    $-19.3$       \\ \hline
\end{tabular}
\end{center}
{\bf Table I.} Comparison of the values for $f_8/f_\pi,$ $f_9/f_\pi $ and
$\theta ,$ calculated in the paper, with the results of the papers referenced.
 \\
 
\section{Comparison with data}
We now wish to compare the values obtained above for the ratios
$f_8/f_\pi ,$ $f_9/f_\pi ,$ and for the $\eta $-$\eta ^{'}$ mixing
angle with available experimental data. We shall first consider in more
detail the well-known $\eta ,\eta ^{'}\rightarrow \gamma \gamma $
decays, for which experimental data are more complete than those for
other processes involving light neutral pseudoscalar mesons, and then
briefly mention the $\eta,\eta ^{'}\rightarrow \pi ^{+}\pi ^{-}\gamma
,$ and $J/\psi \rightarrow \eta \gamma ,\eta ^{ '}\gamma $ decays.

\subsection{$\eta ,\eta ^{'}\rightarrow \gamma \gamma $ decays}
For these processes, the inclusion of both the $\eta $-$\eta ^{'}$ mixing and 
the renormalization of the octet-singlet couplings, which leads to the 
predicted amplitudes
\bqry
F_{\eta \gamma \gamma }(0) & = & \frac{\alpha N_c}{3\sqrt{3}\pi f_\pi }\left(
\frac{f_\pi }{f_8}\cos \theta -2\sqrt{2}\frac{f_\pi }{f_9}\sin \theta \right) 
, \\
F_{\eta ^{'}\gamma \gamma }(0) & = & \frac{\alpha N_c}{3\sqrt{3}\pi f_\pi }
\left( \frac{f_\pi }{f_8}\sin \theta +2\sqrt{2}\frac{f_\pi }{f_9}\cos \theta 
\right) .
\eqry
The values of these amplitudes, as extracted from data on widths,
are~\cite{VH} 
\bqry
F_{\eta \gamma \gamma }(0) & = & 0.024\pm 0.001\;{\rm GeV}^{-1}, \NL
F_{\eta ^{'}\gamma \gamma }(0) & = & 0.031\pm 0.001\;{\rm GeV}^{-1}.
\eqry
Calculation with the help of Eqs. (30),(40),(43)-(46) yields
\bqry
F_{\eta \gamma \gamma }(0) & = & 0.025\pm 0.001\;{\rm GeV}^{-1}, \NL
F_{\eta ^{'}\gamma \gamma }(0) & = & 0.030\pm 0.001\;{\rm GeV}^{-1},
\eqry
in excellent agreement with (47).

\subsection{$\eta,\eta ^{'}\rightarrow \pi ^{+}\pi ^{-}\gamma $ decays}
These, as well as the $P^0\rightarrow \gamma \gamma ,$ processes were 
extensively studied by Venugopal and Holstein \cite{VH} in chiral perturbation
theory. The analysis of experimental data for both of these processes done in 
ref. \cite{VH} yields
$$\frac{f_8}{f_\pi }=1.38\pm 0.22,\;\;\;\frac{f_9}{f_\pi }=1.06\pm 0.03,\;\;\;
\theta =(-22.0\pm 3.3)^o,$$ in good agreement with our Eqs. (40),(43),(44).

\subsection{$J/\psi \rightarrow \eta \gamma ,\eta ^{'}\gamma $ decays}
These processes were studied by Kisselev and Petrov \cite{KP}. The values of
$f_8/f_\pi ,$ $f_9/f_\pi $ and $\theta $ extracted in ref. 
\cite{KP} from the experimentally measured $P^0\rightarrow \gamma \gamma $ and
$J/\psi \rightarrow \eta \gamma ,\eta ^{'}\gamma $ widths, as given in the 
last column of Table I of ref. \cite{KP}, which corresponds to conventional 
mass-mixing angle relations, are
$$\frac{f_8}{f_\pi }=1.12\pm 0.14,\;\;\;\frac{f_9}{f_\pi }=1.04\pm 0.08,\;\;\;
\theta =(-18.9\pm 2.0)^o,$$ again in good agreement with our Eqs. 
(40),(43),(44). 

Thus, the three values of $f_8/f_\pi ,$ $f_9/f_\pi $ and $\theta $ agree with 
experiment (at least, as far as the processes considered above are concerned).

\subsection{Comparison with CELLO and TPC/2$\gamma$ results}
As to the remaining $f_\eta /f_\pi ,$ $f_{\eta ^{'}}/f_\pi $ ratios
also calculated in the paper, the experimental values of them, as
extracted from data by the CELLO \cite{CELLO} and TPC/2$\gamma $
\cite{TPC} collaborations, are, respectively,
\bqry
\frac{f_\eta }{f_\pi } & = & 1.12\pm 0.12, \\
\frac{f_{\eta ^{'}}}{f_\pi } & = & 1.06\pm 0.10,
\eqry
and
\bqry
\frac{f_\eta }{f_\pi } & = & 1.09\pm 0.10, \\
\frac{f_{\eta ^{'}}}{f_\pi } & = & 0.93\pm 0.09.
\eqry
While the value calculated for $f_\eta /f_\pi ,$ Eq. (37), clearly agrees with 
both experimental values (49) and (51), the value calculated for $f_{\eta ^{'}}
/f_\pi ,$ Eq. (38), only marginally agrees with (50), and disagrees with (52)
by almost 3 standard deviations.
 
To clarify the difference from the results referred to above, we recall
that the values of $f_\eta /f_\pi $ and $f_{\eta ^{'}}/f_\pi $ are
extracted by both CELLO and TPC/2$\gamma $ from experimental data on
transition form-factors $T_{\eta (\eta ^{'})}(0,-Q^ 2),$ assuming that
the pole mass $\Lambda _{\eta (\eta ^{'})},$ which parametrizes their
fits to the data, can be identified with $2\pi \sqrt{2}\;\!  f_{\eta
(\eta ^{'})}.$ Then, these pole fits to the data are presumed to join
smoothly, as $Q^2\rightarrow \infty ,$ to the perturbative QCD
predictions for $T_{\eta (\eta ^{'})}(0,-Q^2)$ \cite{BL}, i.e., these
fits would then agree with both the QCD asymptotic form $\sim 1/Q^2$
{\em and} its coefficient.

However, the values of both $f_{\eta }$ and $f_{\eta ^{'}}$ quoted by
the two groups, (in MeV) $(94.0\pm 7.1,$ $89.1\pm 4.9)$ \cite{CELLO},
and $(91.2\pm 5.7,$ $77.8\pm 4.9)$ \cite{TPC}, respectively, are all
close to $M(\rho )/(2 \pi \sqrt{2})\approx 86.5$ MeV, thus indicating a
possible connection with the vector meson dominance interpretation of
$\Lambda _{\eta (\eta ^{'})}\approx M(\rho )$ in the range of $Q^2$
investigated, which could obviate the assumptions referred to above.

On the other hand, as remarked by Klabucar and Kekez \cite{KK}, their own
bound-state calculation, as well as the model independent 
calculation by Gasser and Leutwyler \cite{GL}, testing the Goldberger-Treiman 
relations by Scadron \cite{Sca}, and the calculation by Burden {\it et al.} 
\cite{Burden}, all agree that both $f_\eta $ and $f_{\eta ^{'}}$ should be 
noticeably larger than $f_\pi .$ Our results $f_\eta \sim 1.1f_\pi ,$ $f_{
\eta ^{'}}\sim 1.2f_\pi $ are in agreement with this. 

It therefore seems possible that the discrepancy with other
determinations may be due an inability to extract the values of $f_\eta
$ and $f_{\eta ^{'}}$ from the transition form-factors $T_{\eta (\eta
^{'})}(0,-Q^2)$ Sufficiently accurately, at least in the range of $Q^2$
investigated so far. That this may indeed be the case is indicated by
the experimental value $f_{ \pi ^0}=84.1\pm 2.8$ MeV \cite{CELLO},
extracted by the same method, which is again close to $M(\rho )/2\pi
\sqrt{2}.$ The central value of this $f_{\pi ^ 0},$ 84.1 MeV, is $\sim
10$\% below the well established value given in Eq.  (30), $92.4\pm
0.2$ MeV. Such a large discrepancy cannot be explained by, e.g., small
isospin violation, indicating the possibility that the extracted values
for both $f_\eta $ and $f_{\eta ^{'}}$ may well have been
underestimated also.

\section{Concluding remarks}
As a lagniappe, we note that Eq. (31) may be combined with our
extracted values for $f_9,f_\eta,f_{\eta ^{'}}$ to obtain the value of
$A:$
$$A=0.78\pm 0.12\;{\rm GeV}^2.$$
This is consistent with the usual value 0.73 GeV$^2$ determined by the
trace condition for (14),(16) without $f$'s~\cite{content}.

Let us briefly summarize the findings of this work:

i) We have found that chiral perturbation theory suggests that the
natural object to study is the pseudoscalar mass squared matrix
modified by the inclusion of the squared factors of the pseudoscalar
decay constants, rather than the mass squared matrix itself.

ii) We have shown that this modified mass squared matrix leads to new
Schwinger's quartic mass and Sakurai mass-mixing angle relations for
the pseudoscalar meson nonet.

iii) We have used these new relations to calculate the pseudoscalar
decay constants and mixing angle. We have demonstrated that, except
where questions may be raised regarding the reliability of the
extraction of the relevant quantities from direct experimental data,
the results obtained are in excellent agreement with available data.

We thank Philip R. Page for useful conversations.

\end{document}